**Thermal decay of a metastable state: the influence of the re-scattering on the quasistationary dynamical rate**

M. V. Chushnyakova[1,2,*] and I. I. Gontchar[1,3]

[1]*Physics Department, Omsk State Technical University, Omsk, Russia*
[2]*Institute of Physics and Technology, Tomsk Polytechnic University, Tomsk, Russia*
[3]*Physics and Chemistry Department, Omsk State Transport University, Omsk, Russia*

When a Brownian particle, initially being in the potential well, overcomes the barrier and moves to the absorptive border, it still has a chance to be scattered back to the well by thermal fluctuations. We study this phenomenon carefully modeling numerically the motion of the particle with the Langevin equations. Four potentials which coincide near the well and the barrier but differ in the tail (i.e. beyond the barrier) are considered. It is shown that the potential for which the well and the barrier are described by two smoothly joined parabolas ("the parabolic potential") plays a role of a dividing range for the mutual layout of the quasistationary dynamical rate and the widely used in the literature Kramers rate. Namely, for the potentials with a steeper tails, the Kramers rate $R_{K0}$ underestimates the true quasistationary dynamical rate $R_D$, whereas for the less steep tails opposite holds (inversion of $R_D/R_{K0}$). It is proved that the mutual layout of the values of the $R_D$ for different potentials is explained by the re-scattering of the particles from the potential tail.

*Corresponding author: maria.chushnyakova@gmail.com

**I. INTRODUCTION**

The problem of thermal decay of a metastable (quasistationary) state is typical in modern natural sciences [1-8]. An approximate formula for the rate of this decay taking into account the fluctuation-dissipation character of the process was derived by Kramers in Ref. [9]. We write it using the dimensionless quantities as follows:

$$R_{K0} = \omega_K \left\{ \frac{U_c''}{|U_b''|} \right\}^{1/2} \exp(-\varepsilon). \quad (1)$$

Here

$$\varepsilon = \frac{\widetilde{U}_b}{k_B T}; \quad (2)$$

$T$ is the temperature of the environment; $k_B$ is Boltzmann constant; $\widetilde{U}_b$ is the barrier height (in energy units); $U_c''$ ($U_b''$) is the second derivative of the dimensionless potential energy

$$U = \frac{\widetilde{U}}{m\omega_c^2} \quad (3)$$

with respect to the coordinate at the quasistationary (barrier) point. Here $m$ denotes the mass of Brownian particle (inertia parameter) in proper units; $\omega_c$ is the frequency of oscillations near the bottom of the potential well in units of inverse time. In Eq. (3) $\widetilde{U}$ is the potential energy in energy units.

The multiplier $\omega_K$ reads

$$\omega_K = \left(\frac{\omega_b^2}{\omega_c^2} + \frac{\beta^2}{4}\right)^{1/2} - \frac{\beta}{2}. \tag{4}$$

Here

$$\beta = \frac{\eta}{m\omega_c} \tag{5}$$

is the dimensionless damping coefficient; $\eta$ is the friction coefficient in proper units;

$$\omega_{b(c)} = \sqrt{\frac{|\tilde{U}''_{b(c)}|}{m}}. \tag{6}$$

The dimensionless rate $R_{K0}$ in Eq. (1) is related to the physical rate $\tilde{R}_{K0}$:

$$R_{K0} = \frac{\tilde{R}_{K0}}{\omega_c}. \tag{7}$$

We will refer to the $R_{K0}$ defined by Eq. (1) as to the Zero-order Kramers rate Formula (ZKF).

The conditions of applicability of Eq. (1) can be summarized as follows:

(i) the potential barrier is high enough compare to the thermal energy $k_B T$;

(ii) the absorptive border is far enough from the barrier;

(iii) the quasistationary point is far enough from the barrier;

(iv) the potential is represented well by the portions of parabolas near the quasistationary and barrier points.

The accuracy of the ZKF has got some attention recently [10 – 15]. It was studied by means of comparison with the long time limit of the escape rate obtained using either the stochastic differential equations (the Langevin equations) [10, 11, 13, 15] or the corresponding partial differential equations (the Smoluchowski equation) [12, 14]. This limit is referred to as the Quasistationary Dynamical Rate (QDR) henceforth and denoted as $R_D$.

One sees only the characteristics of the metastable state and of the barrier point in ZKF (Eqs. (1)-(6)). Therefore it could be thought that what is happening to Brownian particles beyond the barrier (during the descent) is not accounted for in ZKF. However this is not true. We showed in [12, 13, 15, 16] that ZKF agrees with the quasistationary dynamical rate only when the absorption point is far enough from the barrier point. In [12, 13, 15, 16] it was shown that for the parabolic potential the two rates (the Kramers rate of Eq. (1) and the QDR) agree within typically 2% for different values of the barrier height, of the temperature, of the barrier curvature, of the location of the absorption point. We consider this agreement to be a proof that in the Kramers formula (1) all the re-scatterings beyond the saddle are accounted for, although implicitly.

In [12 – 15] mostly the influence of the barrier and of the potential well shape on the relation between the ZKF and QDR was studied. Results of these works suggest that the ZKF usually agrees with $R_D$ within 20%. Moreover, in the cases when the disagreement is more significant, $R_{K0}$ is smaller than $R_D$.

When Brownian particle overcomes the potential barrier and escapes from the metastable state, there is still a chance for the particle to be re-scattered back to the well due to thermal fluctuations. This re-scattering can alter the value of $R_D$. In the present work we concentrate on this phenomenon.

The paper is organized as follows. Sec. II is devoted to the description of the model. In Sec. III we compare the Kramers rate with the QDR for the potentials with different tails. In Sec. IV we summarize our results.

**II. THE MODEL**

**A. The scenario**

Our work stems from the nuclear fission problem which was mentioned in the original Kramers paper [9] as one of the examples of thermal decay of a metastable state. This problem involves several degrees of freedom (DOF) [4, 7, 8, 11] and in Ref. [15] we studied the effects of multidimensionality on the accuracy of the Kramers-type approximate formula. However, later we realized that the re-scattering problem is in fact related to the only DOF corresponding to the decay of the metastable state. Therefore in the present work the motion of the Brownian particle is characterized by a single collective coordinate $q$ which is dimensionless. In the case of nuclear fission, $q$ is responsible for the elongation of the fissioning nucleus.

Since we are interested in the back scattering of the particles which already have overcome the barrier, we concentrate on the beyond-barrier shape of the potential and on the location of the absorption point. We consider four potentials presented in Fig. 1. The basic one is constructed of two smoothly joint parabolas ("parabolic potential" $U_P$):

$$U_P(q) = C_0(q - q_c)^2/2 \quad \text{at } q < q_m, \tag{8}$$

$$U_P(q) = U_b - C_0(q - q_b)^2/2 \quad \text{at } q > q_m, \tag{9}$$

$$q_m = (q_b + q_c)/2, \tag{10}$$

$$C_0 = U_b/(q_b - q_c)^2. \tag{11}$$

For the quasistationary and barrier coordinates we use $q_c = 1.00$, $q_b = 1.60$.

Because of thermal fluctuations, the Brownian particle initially located near the metastable state can reach the barrier point with the coordinate $q_b$. The difference between the potential energies at $q_c$ and $q_b$, $U_b - U_c$, is called the barrier height in nuclear fission or the activation energy in chemical reactions. Henceforth we set $U_c = 0$. After reaching the barrier, the particle can return to the quasistationary state due to fluctuations or move further to the absorptive point $q_a$ due to the driving force. The absorptive point in nuclear fission corresponds to the scission point at which the nucleus separates quickly into two fragments.

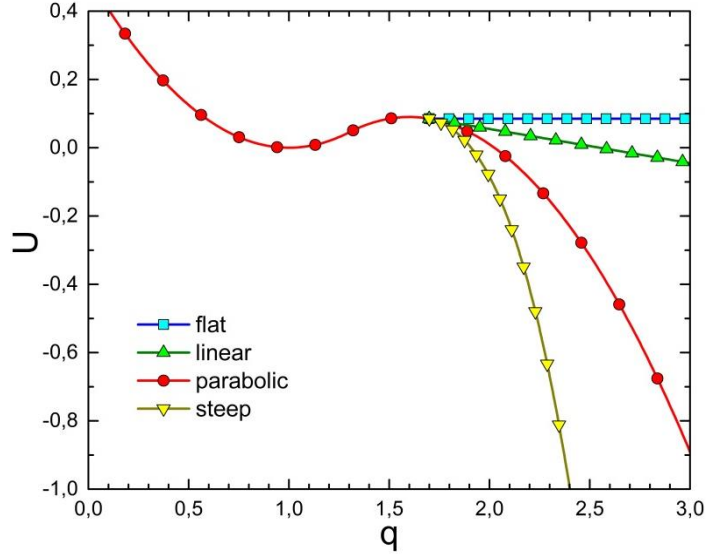

FIG. 1. Four dimenmsionless potentials used in the present work: "flat", "linear", "parabolic", and "steep" (see Eqs. (8)-(14)). $q_c = 1.0$, $q_b = 1.6$, $q_j = 1.7$.

The other three potentials coincide with the parabolic one at $q < q_j$ and differ beyond $q_j$. In Fig. 1 $q_j = 1.7$. From [15, 16] we know that for the parabolic potential the maximum value of $R_D$ is obtained when the absorptive border $q_a$ simply coincides with the barrier. This is equivalent to have a tail of the potential which drops abruptly at $q_a$. As a sample of the potential that is close to this but still not abrupt we use the "steep potential" $U_S$ that reads

$$U_S(q) = U_P(q_j) - C_3(q - q_j)^3/3 \quad \text{at } q > q_j. \tag{12}$$

This potential should result in less back scattering than in the case of the parabolic one.

The "linear potential" $U_L$ is defined as

$$U_L(q) = U_P(q_j) + \left(\frac{dU_P}{dq}\right)_{q_j}(q - q_j) \quad \text{at } q > q_j. \tag{13}$$

It is expected to provide more back scattering than the parabolic one does.

As a limiting case we consider a potential shelf ("flat potential" $U_F$):

$$U_F(q) = U_P(q_j) \quad \text{at } q > q_j \tag{14}$$

which hopefully provides even more back scattering.

## B. Dynamical equations and corresponding decay rates

The time evolution of the dynamical variables of the Brownian particle is described by the stochastic differential equations (the Langevin equations, see Appendix). These equations in discrete form read

$$p^{(n+1)} = p^{(n)} + \Delta p, \tag{15}$$

$$q^{(n+1)} = q^{(n)} + \Delta q, \tag{16}$$

$$\Delta p = -\left\{\beta p + \frac{dU}{dq}\right\}\tau + b\sqrt{\beta U_b \tau/\varepsilon}, \tag{17}$$

$$\Delta q = \frac{p^{(n)} + p^{(n+1)}}{2}\tau. \tag{18}$$

The dimensionless momentum $p$ is related to the physical momentum $\tilde{p}$ as follows

$$p = \frac{\tilde{p}}{m\omega_c}. \tag{19}$$

The superscripts represent two moments of time separated by the time interval $\tilde{\tau} = \tau/\omega_c$, $\tau$ is the dimensionless time step of numerical modeling. In the rhs of Eq. (17) all quantities correspond to the time moment $n\tilde{\tau}$. The amplitude of the random force (last term in Eq. (17)) is related to the temperature (via $\varepsilon$) and to the friction coefficient (via $\beta$) by the fluctuation-dissipation theorem. The random number $b$ entering the random force has a Gaussian distribution with zero average and variance equal to 2.

Eqs. (15)-(18) describe the Markovian process, i.e. the memory effects are not taken into account. These equations are solved numerically by means of the Euler-Maruyama method [18] using random numbers. The solution is actually a sequence of $N_{tot}$ trajectories all terminated not later than at the moment of time $t_D$. Some of those trajectories reach the absorptive point before $t_D$. The dimensionless decay rate is calculated in this algorithm as follows

$$R_a(t) = \frac{1}{N_{tot} - N_{at}} \frac{\Delta N_{at}}{\Delta t}. \tag{20}$$

Here $N_{at}$ is the number of Brownian particles (or stochastic trajectories) which have reached the absorptive point by the time moment $t$, $\Delta N_{at}$ is the number of particles which have reached the absorptive point during the time interval $\Delta t$. Note, that we measure the time in units of $\omega_c^{-1}$. The algorithm for finding $R_D$ is described in detail in [15].

Typical behavior of $R_a(t)$ for the four potentials under consideration is shown in Fig. 2. After a transient stage, the decay rate reaches a quasistationary regime although significant fluctuations are present. Duration of the transient stage depends strongly upon the shape of the potential: the steeper the tail, the shorter the duration stage. Thus care should be taken when choosing the interval for calculating the QDR. Each time we checked whether the results of modeling did not depend upon the time step within the statistical errors. The value of $\tau$ typically was varied from 0.15 up to 0.60.

For the flat potential there is a discontinuity in the force at $q_j$. It is possible to make smooth connection between the parabolic and flat parts of the potential. We made several calculations with such "smooth-flat" potential and found that the values of QDR for the flat and smooth-flat potentials differ not more than by 1-2%. This is typical statistical error of our present calculations.

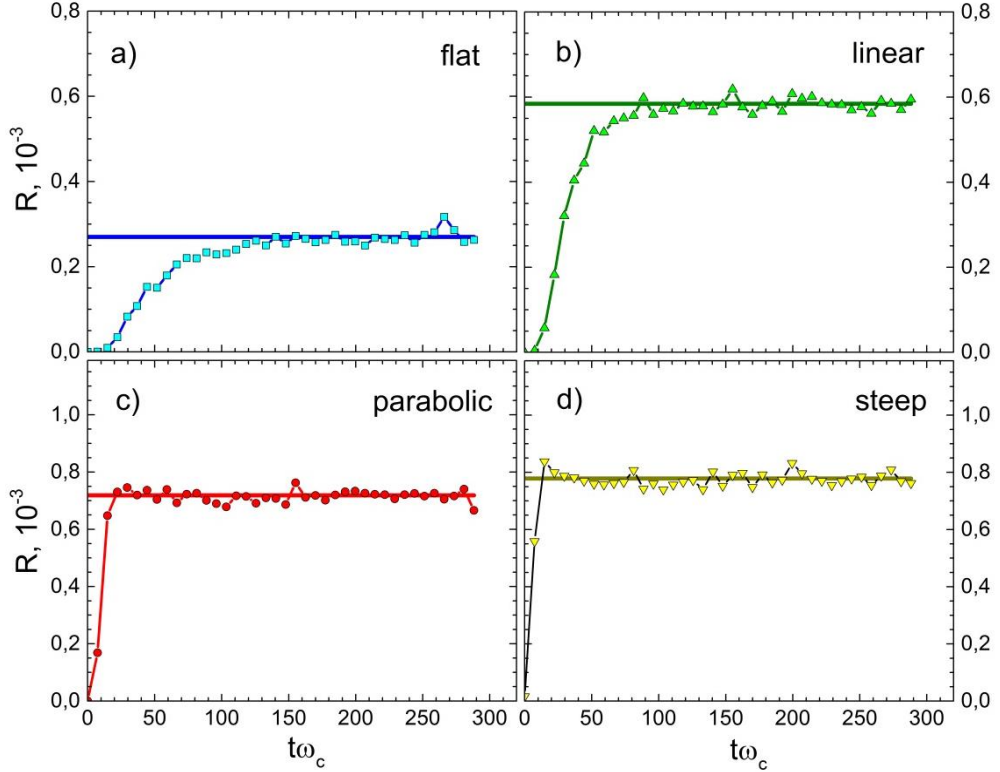

FIG. 2. Typical behaviour of the decay rate $R_a(t)$ for the potentials presented in Fig. 1. The horizontal lines indicate the quasistationary dynamical rates. $\varepsilon = 3.75$, $q_j = 1.7$, $q_a = 2.6$.

**III. RESULTS**

Let us first study what is the role of re-scattering at different values of $\varepsilon$ and how the ZKF measures up against QDR versus $\varepsilon$ when the absorptive point is far enough from the barrier ($q_a = 2.6$) and the junction point where our potentials start to differ is rather close to the barrier ($q_j = 1.7$). Results obtained under these conditions are shown in Fig. 3. Since the ZKF suggests the exponential dependence of $R_{K0}$ upon $\varepsilon$ (see Eq. (1)), we present in Fig. 3a the dependence $R_D(\varepsilon)$ in the logarithmic scale (scattered symbols). One sees that this dependence is exponential indeed for all the potentials whereas the absolute values are somewhat different indicating the influence of the potential tails. The values of $R_D$ for the flat potential are significantly (by factor of 3) below the others. To see clearer the difference between $R_D$ and $R_{K0}$ we display in Fig. 3b the fractional difference

$$\xi_{0D} = R_{K0}/R_D - 1 \qquad (21)$$

with the statistical errors (both in percent) for the parabolic, linear and steep potentials. The curve corresponding to the flat potential lies significantly higher ($\xi_{0D} \sim 200\%$). Recalling the conditions of applicability of the ZKF one realizes that the best agreement between $R_D$ and $R_{K0}$ for the parabolic potential is to be expected. A less expected feature of $\xi_{0D}$ for this potential is that it does not increase in absolute value as $\varepsilon$ decreases. This effect was discussed in detail in Ref. [13]. For the case of the other two potentials we see that for the steep one the ZKF underestimates the dynamical rate by some 10% ($\xi_{0D} < 0$) whereas for the linear potential the ZKF overestimates the rate by approximately 20% ($\xi_{0D} > 0$).

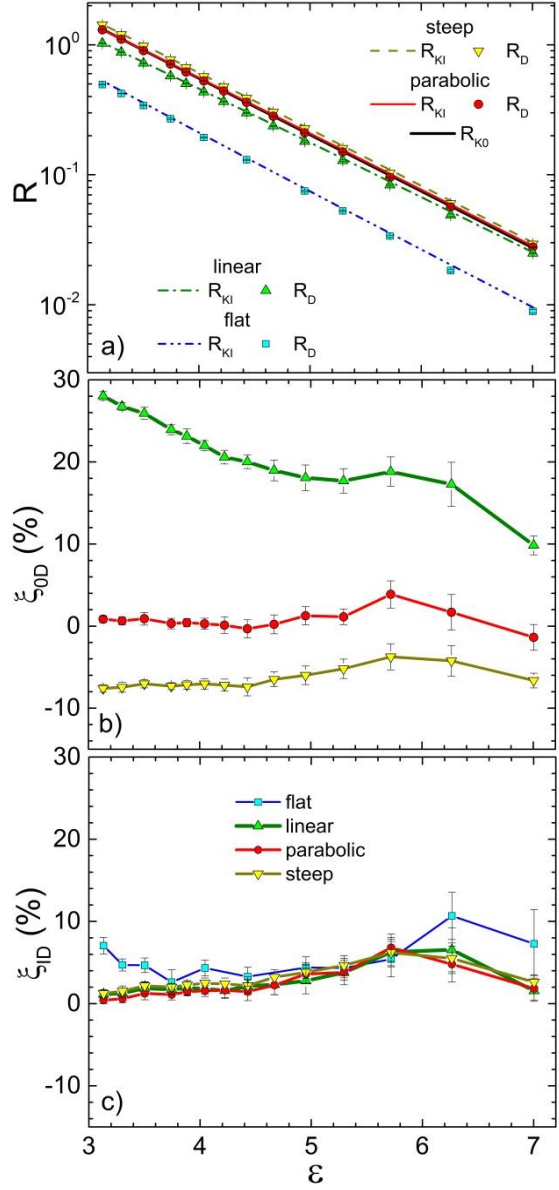

FIG. 3. (a) Quasistationary dynamical rates (symbols) and integral Kramers rates (lines) versus $\varepsilon$ for the potentials presented in Fig. 1. Zero-order Kramers rate is shown as well but it is indistinguishable from $R_{KI}$ for the parabolic potential. (b) Fractional difference $\xi_{0D}$ defined by Eq. (21), (c) fractional difference $\xi_{ID}$ defined by Eq. (24). $q_j = 1.7$, $q_a = 2.6$.

Let us now come back to Fig. 3a and discuss the lines represented there. These lines correspond to the so-called Integral Kramers Formula (IKF). This formula for the decay rate was discussed in detail in Refs. [12, 13, 17]. In fact it was implied (but not written explicitly) in the original Kramers paper [9]. The dimensionless IKF reads

$$R_{KI} = \frac{U_b}{\beta \varepsilon} \left\{ \int_{-\infty}^{q_b} \exp\left[-\frac{\tilde{U}(x)}{k_B T}\right] dx \int_{q_c}^{q_a} \exp\left[\frac{\tilde{U}(y)}{k_B T}\right] dy \right\}^{-1}. \qquad (22)$$

In Eq. (22) the integral from $-\infty$ up to $q_b$ represents the population near the quasistationary point, whereas the inverse integral from $q_c$ down to $q_a$ is proportional to the flux over the potential barrier. The formulas leading us to (22) are scattered in pages 290-293 of [9] and often not numbered.

Formally, Eq. (22) is valid in the case of large friction when the motion is overdamped:

$$\frac{\beta^2}{4} \gg 1. \qquad (23)$$

In our case $\beta^2/4 = 6.43$, thus inequality (23) approximately holds. It was shown in [12, 13] that the IKF provides better approximation to QDR in the cases when the potential deviates from the parabolic shape near the barrier and the quasistationary point. In Fig. 3a one sees that it is true in our case too. In Fig. 3c we quantify the relation between the IKF and QDR showing the fractional difference

$$\xi_{ID} = R_{KI}/R_D - 1 \qquad (24)$$

in the same scale as in Fig. 3b. In Fig. 3c we see that the values of $\xi_{ID}$ for different potentials are very close to each other. It means that in the considered case the IKF works very well too. Comparing Figs. 3b and 3c one notices that the parabolic potential (circles) is the only one for which the values of $\xi_{ID}$ and $\xi_{0D}$ are close to each other. For the linear potential (triangles up) $\xi_{ID}$ lies within 10% whereas $\xi_{0D}$ exceeds 25%. It is even possible to present $\xi_{ID}$ for the flat potential (squares) in the same figure (Fig. 3c) whereas $\xi_{0D}$ estimated from Fig. 3a exceeds 200%.

In order to prove that this is the backscattering which results to $R_D$(flat) < $R_D$(linear) < $R_D$(parabolic) < $R_D$(steep) we register the re-scattered particles. There are two features of a particle to be registered as the backscattered one: (i) its coordinate at least once takes a value larger than $q_b$; (ii) at the end of calculation the coordinate of this particle is smaller than $q_b$. Results of this registration are presented in Table I for two values of $\varepsilon$. One sees that the number of re-scattered particles is very significant in comparison with the number of absorbed particles.

TABLE I. The QDR and the number of re-scattered particles evaluated numerically, $N_{rs\,num}$, and analytically, $N_{rs\,an}$, over the number of absorbed particles, $N_a$, for four potentials. $\varepsilon = 4.43$ and $3.74$; $q_a = 2.6$; $q_j = 1.7$; $N_{tot} = 5 \cdot 10^5$; $t_D = 300$ and $600$. The adjusting parameter $\alpha$ is explained in the text.

|  | $\varepsilon = 4.43$; $R_{K0} = 0.3594\,10^{-3}$; $\alpha = 1.560$ | | | $\varepsilon = 3.74$; $R_{K0} = 0.7137\,10^{-3}$; $\alpha = 1.506$ | | |
|---|---|---|---|---|---|---|
|  | $t_D = 300$ | | | $t_D = 300$ | | |
|  | $R_D, 10^{-3}$ | $N_{rs\,num}/N_a$ | $N_{rs\,an}/N_a$ | $R_D, 10^{-3}$ | $N_{rs\,num}/N_a$ | $N_{rs\,an}/N_a$ |
| flat | 0.1306 | 3.849 | 4.276 | 0.2697 | 3.688 | 3.830 |
| linear | 0.2985 | 0.998 | 1.043 | 0.5842 | 1.048 | 1.039 |
| parabolic | 0.3587 | 0.620 | === | 0.7184 | 0.611 | === |
| steep | 0.3850 | 0.504 | 0.495 | 0.7786 | 0.487 | 0.441 |
|  | $t_D = 600$ | | | $t_D = 600$ | | |
| flat | 0.1284 | 3.281 | 3.925 | 0.2644 | 2.637 | 3.663 |
| linear | 0.3001 | 0.865 | 1.011 | 0.5773 | 0.774 | 1.076 |
| parabolic | 0.3553 | 0.571 | 0.654 | 0.7076 | 0.482 | 0.648 |
| steep | 0.3833 | 0.466 | 0.524 | 0.7677 | 0.387 | 0.504 |

In addition to the number of registered re-scattered particles during the numerical modeling $N_{rs\,num}$ we show in this Table the QDR, the number of escaped (absorbed) particles $N_a$, and the number of re-scattered particles $N_{rs\,an}$ which we calculated analytically from the following consideration.

Neglecting the transient stage, one can estimate the number of non-absorbed particles from the radioactive decay law

$$N_{tot} - N_a = N_{tot}\exp(-R_D t_D). \tag{25}$$

In all our calculations $R_D t_D \ll 1$ therefore approximately

$$\frac{N_a}{N_{tot}} = R_D t_D. \tag{26}$$

It seems reasonable to accept that the number of absorbed particles is just the difference between the number of the particles which have overcome the barrier $N_b$ and the number of re-scattered particles $N_{rs}$:

$$N_a = N_b - N_{rs}. \tag{27}$$

The number of the particles that have not overcome the barrier should follow the same radioactive decay law but with the rate which is between $R_{K0}$ and $2R_{K0}$:

$$N_{tot} - N_b = N_{tot}\exp(-\alpha R_{K0} t_D). \tag{28}$$

Here $\alpha$ is still unknown factor. Since $\alpha R_{K0} t_D \ll 1$, Eq.(28) results in

$$N_b/N_{tot} = \alpha R_{K0} t_D. \tag{29}$$

Combining now Eqs. (26), (27), (29) and excluding $N_a$ and $N_b$ we arrive at

$$N_{rs} = N_{tot} t_D (\alpha R_{K0} - R_D). \tag{30}$$

We now can estimate $N_{rs}$ analytically ($N_{rs\,an}$) if we know $\alpha$. To find it we first apply Eq. (30) for the parabolic potential using known numerical value of $N_{rs}$ ($N_{rs\,num}$).

In Table I we see that this algorithm provides rather reasonable results. First, $\alpha$ lies between 1 and 2 as we expected. Second, for two rather different values of $\varepsilon$ the values of $\alpha$ are very close. Third, the number of re-scattered particles obtained analytically is close to that resulting from numerical modeling. No exact equality between $N_{rs\,an}$ and $N_{rs\,num}$ is to be expected because in our derivation we neglect the transient stage and double re-scattering. A propos, in Fig. 2 we see that the longest transient stage corresponds to the flat potential for which the agreement between $N_{rs\,an}$ and $N_{rs\,num}$ is the worst. In our opinion, results of Table I prove that the inversion of $R_D/R_{K0}$ as the potential goes over from the steep to the flat one is solely due to re-scattering.

All the results above are obtained when the absorptive point is far enough from the barrier ($q_a = 2.6$) and the junction point is rather close to the barrier ($q_j = 1.7$). Let us now see how the ZKF and IKF measure up against QDR when the absorptive border moves closer and further to the saddle point. The fixed parameters for these calculations are $\varepsilon=3.175$ and $q_j = 1.70$. Results are shown in Fig. 4. Here we see that for the potentials which are "softer" than the parabolic (i.e. decreasing slower after the junction point) the ZKF significantly overestimates the true dynamical decay rate. Thus the mutual layout of $R_{K0}$ and $R_D$ definitely inverses as one switches over from a potential which is steeper than parabolic to the one which is flatter. Moreover contrary to the cases of the linear, parabolic, and steep potentials, $R_D$ for the flat potential does not reach any plateau with the increase of $q_a$. The IKF reproduces the numerical QDR nicely at all values of $q_a$.

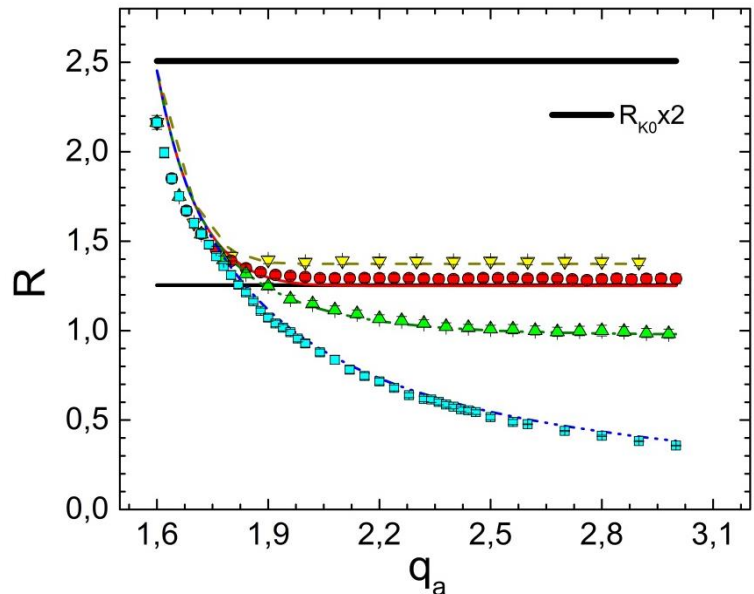

FIG. 4. The quasistationary dynamical rates (symbols) and the rates calculated according to the integral Kramers formula (lines) versus the absorption point coordinate for four potentials presented in Fig. 1. The notations are the same as in Fig. 3a. Thick solid line in the upper part corresponds to the doubled zero-order Kramers rate. $\varepsilon = 3.17$, $q_j = 1.7$.

Finally in Fig. 5 we show the evolution of the rates with the increase of the junction point coordinate $q_j$. As it might be expected, all the rates converge to the ZKF as $q_j$ increases.

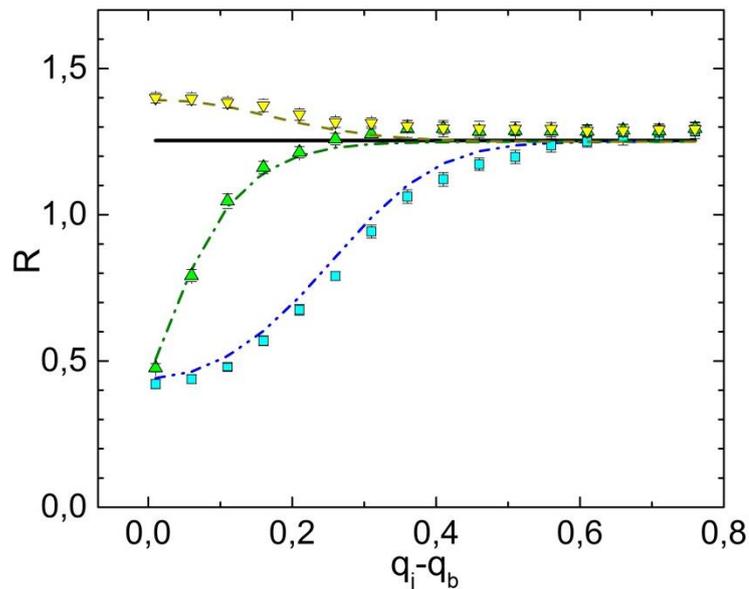

FIG. 5. The same as in Fig. 4 but versus the difference $q_j - q_b$. Results for the parabolic potential are not shown because $q_j$ is not applicable in this case. $\varepsilon = 3.17$, $q_a=2.6$.

**IV. CONCLUSIONS**

We have studied the effect of the re-scattering of the Brownian particles in their wandering out of the metastable state beyond the barrier. Four potentials have been considered which coincide near the potential well and the barrier but differ beyond the barrier. The conclusions can be summarized as follows:

(i) the steep potential results in the quasistationary dynamical rate $R_D$ that is larger than the Kramers rate $R_{K0}$ of Eq. (1), whereas for the linear and flat potentials opposite holds (inversion of $R_D/R_{K0}$);

(ii) the $R_{K0}$ disagrees with $R_D$ significantly for all but parabolic potential;

(iii) we derived a formula (Eq. (30)) which allows to estimate analytically the number of re-scattered particles;

(iv) the mutual layout of the values of the $R_D$ for different potentials is explained by the re-scattering of the particles from beyond the barrier.

**APPENDIX**

For the one-dimensional case the Langevin equations in the differential form read

$$\begin{cases} dq = \tilde{p}m^{-1}d\tilde{t}, \\ d\tilde{p} = -\left(\frac{\eta}{m}\tilde{p} + \frac{d\tilde{U}}{dq}\right)d\tilde{t} + \sqrt{2\eta\theta}d\widetilde{W}. \end{cases} \quad (31)$$

Here the coordinate $q$ is dimensionless; all other quantities have physical dimensions.

In Eq. (31) $\widetilde{W}$ is the Wiener process whose increment $d\widetilde{W}$ possesses the normal distribution with the variance $d\tilde{t}$; $\theta$ is the thermal energy. For example, in nuclear physics $\theta$ is equal to the temperature $T$ measured in MeV, in chemical or molecular applications $\theta = k_B T$ ($k_B$ is the Boltzmann constant). The time interval during which the momentum changes by $d\tilde{p}$ and the coordinate changes by $dq$ is denoted as $d\tilde{t}$.

The values of all these quantities significantly depend upon the particular physical problem which the Langevin equations are used for. Nevertheless the main features of thermal decay of a metastable state are common. Therefore it is useful to convert these equations into dimensionless form to exclude such particularity and to emphasize commonness.

The relations between the quantities $\tilde{p}, \eta, \tilde{U}, \theta$ (or the same $k_B T$) with the dimensionless ones $p, \beta, U, \varepsilon$ are presented in Eqs. (19), (5), (3), (2), respectively. Since for the dimensionless time we have

$$t = \tilde{t}\omega_c \quad (32)$$

($\omega_c$ is defined in (6)), for the dimensionless increment of the Wiener process the equality

$$dW = d\widetilde{W}\omega_c^{1/2} \quad (33)$$

should hold.

Appling all these notations we come to the dimensionless Langevin equations:

$$\begin{cases} dq = pdt, \\ dp = -\left(\beta p + \dfrac{dU}{dq}\right)dt + \sqrt{2\beta \dfrac{U_b}{\varepsilon}}\,dW. \end{cases} \quad (34)$$

Equations (15)-(18) represent the discretized version of Eqs. (34).